\documentclass[conference,a4paper,10pt]{IEEEtran}
\IEEEoverridecommandlockouts

\makeatletter
\AtBeginDocument
 {
   \def\ltx@label#1{\cref@label{#1}}
   \def\label@in@display@noarg#1{\cref@old@label@in@display{#1}}
\def\label@in@mmeasure@noarg#1{%
    \begingroup%
      \measuring@false%
      \cref@old@label@in@display{#1}
    \endgroup}%
 } %
\makeatother

\usepackage{amsmath,graphicx,amssymb,mathrsfs,amsfonts}

\usepackage{algorithm}
\usepackage[noend]{algpseudocode}
\usepackage{algorithmicx}

\usepackage{csquotes}

\usepackage{hyperref}
\usepackage[nameinlink,capitalize,noabbrev]{cleveref}
\usepackage[nolist,nohyperlinks]{acronym}
\usepackage{xcolor}
\usepackage{amsmath}
\usepackage{amssymb}
\usepackage{amsthm}
\usepackage{tikz}
\usepackage{tikz-cd}
\usepackage{fancyhdr}

\usepackage{cite}
\usepackage{textcomp}

\def\BibTeX{{\rm B\kern-.05em{\sc i\kern-.025em b}\kern-.08em
    T\kern-.1667em\lower.7ex\hbox{E}\kern-.125emX}}

\newcommand{\norm}[1]{\left\Vert#1\right\Vert}

\DeclareMathOperator{\grad}{grad}
\DeclareMathOperator{\curl}{curl}
\DeclareMathOperator{\harm}{harm}

\newcommand{\Grad}[1]{\ensuremath \grad_\cc(\mathbf{#1})}
\newcommand{\Curl}[1]{\ensuremath \curl_\cc(\mathbf{#1})}
\newcommand{\Harm}[1]{\ensuremath \harm_\cc(\mathbf{#1})}

\theoremstyle{definition}
\newtheorem{definition}[equation]{Definition}
\newtheorem*{definition*}{Definition}

\theoremstyle{remark}

\newtheorem{remark}[equation]{Remark}

\DeclareMathOperator{\Ima}{Im}

\newcommand{\cc}{\mathcal{C}}
\newcommand{\CC}{{CC}}

\newcommand{\subheading}[1]{\noindent \textbf{#1}}

\acrodef{MFCI}{Matrix-Factorization-Cell-Inference}
\acrodef{SVD}{Singular Value Decomposition}
\acrodef{SPH}{Spanning Tree Heuristic}
\acrodef{ICA}{Independent Component Analysis}

\begin{document}

\title{Faster Inference of Cell Complexes from Flows via Matrix Factorization%
	\thanks{Funded by the European Union (ERC, HIGH-HOPeS, 101039827). Views and opinions expressed	are however those of the authors only and do not necessarily reflect those of the European Union or the European Research Council Executive Agency. Neither the European Union nor the granting authority can be held responsible for them.

		Parameter search was performed with computing resources granted by RWTH Aachen University under project \emph{thes1725}.\looseness=-1

		Evaluation code:	 \url{https://github.com/til-spreuer/mfci-workflow}}%
}

\author{\IEEEauthorblockN{1\textsuperscript{st} Til Spreuer}
	\IEEEauthorblockA{%
		\textit{RWTH Aachen University}\\
		Aachen, Germany \\
		til.spreuer@rwth-aachen.de}
	\and
	\IEEEauthorblockN{2\textsuperscript{nd} Josef Hoppe}
	\IEEEauthorblockA{%
		\textit{RWTH Aachen University}\\
		Aachen, Germany \\
		orcid.org/0000-0003-4383-7049}
	\and
	\IEEEauthorblockN{3\textsuperscript{rd} Michael T.\ Schaub}
	\IEEEauthorblockA{%
		\textit{RWTH Aachen University}\\
		Aachen, Germany \\
		orcid.org/0000-0003-2426-6404}
}

\maketitle
\thispagestyle{fancy}
\fancyfoot[L]{Accepted at EUSIPCO 2025, Palermo, Italy}
\pagestyle{plain}
\bstctlcite{IEEEexample:BSTcontrol}

\begin{abstract}
	We consider the following inference problem: Given a set of edge-flow signals observed on a graph, lift the graph to a cell complex, such that the observed edge-flow signals can be represented as a sparse combination of gradient and  curl flows on the cell complex.
Specifically, we aim to augment the observed graph by a set of 2-cells (polygons encircled by closed, non-intersecting paths), such that the eigenvectors of the Hodge Laplacian of the associated cell complex provide a sparse, interpretable representation of the observed edge flows on the graph. 
As it has been shown that the general problem is NP-hard in prior work, we here develop a novel matrix-factorization-based heuristic to solve the problem.
Using computational experiments, we demonstrate that our new approach is significantly less computationally expensive than prior heuristics, while achieving only marginally worse performance in most settings. In fact, we find that for specifically noisy settings, our new approach outperforms the previous state of the art in both solution quality and computational speed.
\end{abstract}

\section{Introduction}

Graphs have become a prevalent abstraction in data science due to their ability to model many real-life systems \cite{mulder2018network}.
Graph signal processing (GSP) \cite{shuman2013emerging,ortega2018graph} enables signal processing for signals that are defined on such graphs, such as temperatures measured at different locations or neuron activity in different parts of the brain. 
However, in many applications, the recorded data represents flows, e.g.\ of people~\cite{lee2011study}, traffic \cite{gh-transportation-networks}, or money \cite{iori2008network}.
Such flows are often more naturally represented with an (oriented) signal on the edges of a graph.
To process such edge signals, recent extensions of GSP towards topological signal processing (TSP) \cite{barbarossa2020topological2,sardellitti2021topological,schaub2021signal,schaub2018flow,schaub2022signal,roddenberry2022signal} utilize simplicial complexes or cell complexes, and their associated Hodge Laplacians.

These Hodge Laplacian operators, a generalization of the graph Laplacian to higher dimensions, are a central pillar of TSP, and can be used in lieu of the graph Laplacian as shift operators in signal processing tasks for signals defined on the edges or higher-dimensional cells of a complex.
Importantly, the 1-Hodge Laplacian induces a decomposition of the space of edge flows into gradient, curl, and harmonic flows \cite{lim2019hodgelaplaciansgraphs, Schaub_2020}.
Similarly to the graph Laplacian in GSP, the eigenvalues and corresponding eigenspaces of this Hodge Laplacian enable the construction of low- and high-pass filters, denoising, signal compression, and other classical signal processing tasks \cite{schaub2021signal}.

However, the higher-order cell structure of the complexes used to construct these Hodge-Laplacians is typically not readily available.
Specifically, when considering edge-flow data, commonly only the underlying graph structure and the associated edge flows are observable.
Therefore, analogously to inferring a graph structure from data on the nodes \cite{ortega2018graph}, different methods to infer simplicial complex from edge data on a graph have been proposed \cite{sardellitti2021topological,battiloro2023latent}.
The general problem of inferring a cell complex from such data, i.e., finding an optimal set of $2$-cells (polygons)
such that the edge-flows can be represented by a sparse set of eigenvectors of the Hodge Laplacian, was introduced in~\cite{hoppe2023representing}.
It was shown that this problem is NP-hard, and thus \cite{hoppe2023representing} introduced a heuristic algorithm to solve it.

\subheading{Contribution.} In this paper, we introduce an alternative approach to solving the cell inference problem using matrix decompositions.
As we show, our novel approach is significantly faster and incurs only a marginally higher approximation error than the originally proposed heuristic.
In our experiments, this trade-off is particularly advantageous on larger networks, when inferring a large absolute number of 2-cells, and in the presence of high noise.

\section{Background}
This introduction to $2$-dimensional cell complexes is inspired from \cite{hoppe2025cellcomplexes}.

We consider $2$-dimensional complexes constructed by augmenting simple undirected graphs as follows.
Consider a graph ${\mathcal{G}=(\mathcal{V}, \mathcal{E})}$ with a set of vertices $\mathcal{V} = \{v_1, \ldots, v_n\}$ and a set of edges by $\mathcal{E} = \{e_1, \ldots, e_m\}$.
Each edge $e_k$ consists of a pair of vertices $e = (v_{i}, v_{j})$, with arbitrary but fixed order.
We call $v_i$ the \emph{source} and $v_j$ the \emph{target} of $e_k$.
We encode the structure of the graph in an incidence matrix $\mathbf{B}_1 \in \{0, \pm 1\}^{n \times m}$.
For every edge $e_k$, oriented from $v_i$ to $v_j$, we have $(\mathbf{B}_1)_{i,k} =1$ and $(\mathbf{B}_1)_{j,k}=-1$, and $(\mathbf{B}_1)_{-,k} = 0$ otherwise.

\begin{definition}[Simple cell complexes of dimension $2$]	\label{def:2dCellComplexes}
	An abstract regular cell complex $\cc$ of dimension $2$ consists of a graph $\mathcal{G}$, and a non-empty ordered set of $2$-cells (polygons) $\cc_2$, encoded via a boundary matrix $\mathbf{B}_2\in \{0,\pm 1\}^{m \times |\cc_2|}$.
	Specifically for every $2$-cell $\theta \in \cc_2$, we have a set of edges $\{e_1, \ldots, e_m\}$ corresponding to a simple cycle in $\mathcal G$ which forms the boundary of $\theta$.
	Each column of $\mathbf{B}_2$ corresponds to such a cycle and the entries of the associated edges are set to $\pm 1$ such that $\mathbf{B}_1\mathbf{B}_2=0$.
\end{definition}

\begin{remark}[Orientation]
	Note that the above construction amounts to equipping each edge and each polygon with a reference orientation.
	This built-in notion of \emph{orientation} allows for a natural representation of physical data like flows via positive or negative values (with or opposite to the orientation).
\end{remark}

Oriented signals on \CC{}s can be represented as \emph{chains}:

\begin{definition}[Signal (or chain) space of a cell complex]
	\label{def:SignalSpace}
	Given a cell complex $\cc$, we denote by $C_k=\mathbb{R}^{|\cc_k|}$ the $k$-th signal space of $\cc$ and obtain an associated sequence of signal spaces
	\[
		\begin{tikzcd}
			C_0&\ar[l,"\mathbf{B}_1"]C_1&\ar[l,"\mathbf{B}_2"]C_2
		\end{tikzcd}
	\]
\end{definition}
The boundary matrices $\mathbf{B}_k$ are linear maps between the signal spaces $C_k$.
In this paper, we analyze flows from the signal space on edges $C_1$.

\subheading{Hodge Laplacian and Hodge decomposition.}
The Hodge Laplacian $\mathbf{L}_k = \mathbf{B}_k^\top \mathbf{B}_k + \mathbf{B}_{k+1}\mathbf{B}_{k+1}^\top$ is a generalization of the Graph Laplacian \cite{lim2019hodgelaplaciansgraphs,Schaub_2020,discrete_calculus}.
The Hodge decomposition  divides the space of edge flows $C_1$ into three eigenspaces associate to $\mathbf{L}_1$: The \emph{gradient} space $\Ima \mathbf{B}_1^\top$, the \emph{curl} space $\Ima \mathbf{B}_2$, and the \emph{harmonic} space $\ker \mathbf{B}_1^\top \cap \ker \mathbf{B}_2$.
Intuitively, the gradient space consists of flows based on the difference between potentials on the nodes and the curl space consists of flows around the boundaries of $2$-cells.

For a flow \(\mathbf{f} \in C_1\) on a cell complex \(\cc\), we define
the gradient flow \(\Grad{f} = \mathbf{B}_1^\top(\mathbf{B}_1^\top)^\dagger \mathbf{f}\),~curl flow \(\Curl{f} = \mathbf{B}_2(\mathbf{B}_2)^\dagger\mathbf{f}\), and
harmonic flow \({\Harm{f} = (\mathbf{I} - \mathbf{L}_1 (\mathbf{L}_1)^\dagger )\mathbf{f}}\).
\((\cdot)^\dagger\) denotes the Moore-Penrose pseudoinverse.

\section{The Cell Inference Problem}

Given a graph and a set of observed flows, the cell inference problem is to find a sparse set of $2$-cells that minimize the projection of the flows into the harmonic space when added. We denote the given graph by \(\mathcal{G} = (\mathcal{V}, \mathcal{E})\) and
the \(s \in \mathbb{N}\) sampled edge flows by \(\mathbf{f}_i \in C_1\)
for \(i \in \{1, \ldots, s\}\).
In the following, we use $\mathbf{F}$ to denote the matrix of all flows \(\mathbf{F} = \left[\mathbf{f}_1, \ldots, \mathbf{f}_s \right]\).\looseness=-1

More formally, the problem can be modeled as an optimization problem: find a $2$-dimensional cell complex $\cc$ which minimizes the harmonic projection of the flows $\mathbf{F}$ onto the harmonic space of $\cc$.:
\begin{align}\label{eq:loss}
	 \mathcal{L}(\cc,\mathbf{F}) = \|\text{harm}_{\cc}(\mathbf{F})\|_F = \left({\sum_{i=1}^{s}\left|\left|\text{harm}_{\cc}(\mathbf f_i)\right|\right|_2^2}\right)^{1/2} 
\end{align}
As minimizing this loss without constraints on the number of cells would result in a $\cc$ with an empty harmonic space, we formulate the minimization problem such that the number of 2-cells is constrained:
\begin{equation}\label{eq:problem}
	\min_{\cc} \mathcal{L}(\cc,\mathbf{F}) \quad \text{s.t.}
	 \quad \cc \text{ has $\mathcal{G}$ as $1$-skeleton} \text{ and }\left|\cc_2\right| \leq k
\end{equation}
Without loss of generality, we will in the following assume that the flows $\mathbf{F}$ are gradient free, as any gradient component will not alter the optimal $\mathcal{C}$ in our optimization problem.

\section{Methods}

\begin{figure*}[t]
	\centering
	\includegraphics[width=\linewidth]{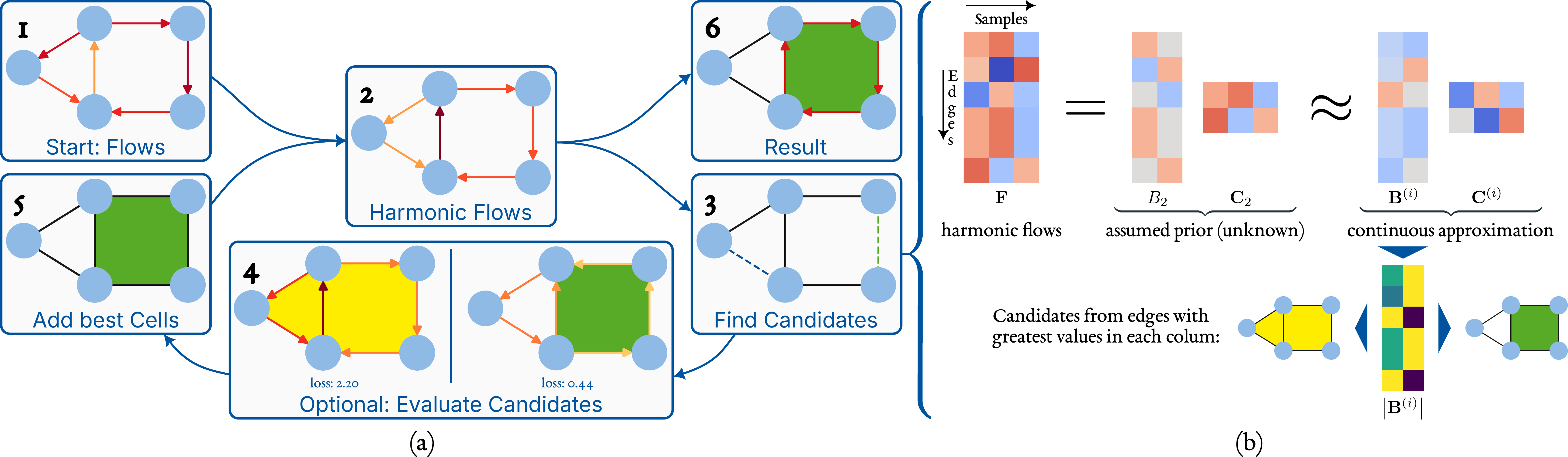}
	\vspace*{-20pt}
	\caption{Overview of the complete inference approach. \textbf{(a)} shows an overview of our Cell Inference Algorithm, adapted and modified from \cite{hoppe2023representing}. Like the original algorithm, our matrix-based alternative takes a graph with flows on its edges as an input (1) and iteratively adds $2$-cells. Both approaches project the flows into the harmonic space (2). Our approach introduces a matrix-factorization-based heuristic for finding candidates (3) that also makes the individual evaluation of the candidates (4) optional in practice. Furthermore, our approach can add multiple cells (instead of one) in each iteration (5). \textbf{(b)} shows the concept of the novel candidate heuristic in \ac{MFCI}. The harmonic flows are decomposed into two matrices, resembling a relaxed version of flows generated by a boundary matrix. Thus, the left matrix can be discretized to one cell candidate per column. Pseudocode of steps 2-5 is given in \cref{alg:matfact_cs}}\label{fig:overview}
\end{figure*}

As proposed in~\cite{hoppe2023representing}, we can proceed in an iterative fashion to solve the cell inference problem.
We start with (i) a cell complex \(\cc^{(0)}\) equivalent to \(\mathcal{G}\), (ii) the number of cells to add per iteration $l'$, and (iii) the number $l$ of cell candidates to consider, with $l' \leq l $.
In each iteration \(i\), we add \(l'\) new 2-cells \(\theta^{(i)}_{1}, ..., \theta^{(i)}_{l'}\) until the desired amount of 2-cells are inferred:
\begin{equation}\label{eq:iterative-update}
	\cc^{(i)} \gets \cc^{(i-1)} \cup \{ \theta^{(i)}_j : j \in [1, \ldots, l'] \}
\end{equation}

To operationalize this algorithmic idea, we have to address two questions.
First, how do we obtain possible cell candidates \(\theta^{(i)}_{1}, ..., \theta^{(i)}_{l}\) in each iteration?
Second, how do we evaluate the candidates in each iteration  and select the best?

In the original work~\cite{hoppe2023representing}, a \ac{SPH} was proposed that constructs candidate 2-cells from one or multiple constructed spanning trees.
The candidate cells that (when added to the cell complex) maximally decrease the loss are then selected greedily.
This requires computing several harmonic projections of an (augmented) cell complex per iteration, which is the computationally most expensive part of the algorithm (even though efficient algorithms like LSMR~\cite{lsmr} can be used to compute the projections).

There is also a more subtle issue associated with the \ac{SPH} approach, that leads to many iterations required to infer appropriate cells.
When using spanning trees to generate candidate cells, each candidate cell contains exactly one edge outside the spanning tree, and at least two edges from the tree.
Importantly, candidate cells that significantly lower the loss are likely to share edges in the spanning tree.
Adding one of those cells to the complex will account for the flow on all the shared spanning tree edges.
The other candidates are thus unlikely to decrease the loss significantly, as they share many edges with the already added cell.
Hence, in \cite{hoppe2023representing}, only \emph{one} cell is added in each iteration.

\subsection*{Matrix-factorization-based approach}\label{sec:methodology}
In the following we present a novel, modular framework called \acf{MFCI} (cf. \cref{fig:overview}, \cref{alg:matfact_cs}) to solve the cell inference problem.
For this, we interpret~\cref{eq:problem} as a problem to find a matrix $\mathbf{B}_2$ that minimizes:
\begin{equation}\label{eq:problem-variant-matfact}
	\min_{\mathbf{B}_2, \mathbf{C}} \norm{\mathbf{F}-\mathbf{B}_2 \mathbf{C}}
	\quad\text{s.t.}\quad \mathbf{B}_2 \in \mathcal{B}_2^{(k')},~
	\mathbf{C} \in \mathbb{R}^{k' \times s}
\end{equation}
where \(\mathcal{B}_2^{(k')}\) is the set of valid edge-to-cell boundary matrices of cell complexes with \(k' \leq k\) 2-cells and a 1-skeleton given by \(\mathcal{G}\); $\mathbf{C}$ is a matrix of real coefficients.
Given a matrix \(\mathbf{B}_2\), an optimal \(\mathbf{C}\) can be obtained by solving a least squares problem.

\begin{algorithm}
	\caption{One Iteration of Candidate Search for \ac{MFCI}}\label{alg:matfact_cs}
	\textbf{Input:} cell complex \CC, the amount of desired 2-cells \(l \in \mathbb{N}_+\)\\
	\textbf{Output:} a set of \(k\) 2-cells
	\begin{algorithmic}[1]
		\State Compute the matrix factorization \(\mathbf{B \cdot C} \approx \harm_\CC(\mathbf{F})\)
		\State Extract the most promising row vectors \(b_1, \ldots, b_l\) from \(\mathbf{B}\)
		\State Discretize \(b_1, \ldots, b_l\) to 2-cells \(\theta_1, \ldots, \theta_l\)
		\State \textbf{Return: } \(\{\theta_1, \ldots, \theta_l\}\)
	\end{algorithmic}
\end{algorithm}

Our key idea now is to first find a low-rank approximation of the flows \(\mathbf{F} \approx \mathbf{B} \cdot \mathbf{C}\) using a matrix factorization, without considering the constraints on $\mathbf{B}$; and then discretize the columns of $\mathbf{B}$ to correspond to valid boundary vectors of 2-cells.
This idea can be iteratively applied several times, adding multiple cells in each iteration.
To create a practical algorithm from this idea we need to choose a suitable matrix factorization method, a heuristic to find cell candidates from this factorization, and a procedure to evaluate and select the best candidates.
As a secondary consideration, we also consider an efficient approximation for calculating the harmonic flows.

\subheading{Calculating (approximate) harmonic flows.} In each step $i$, \ac{SPH} calculates the harmonic flows \({\mathbf{H}^{(i)} := \harm_{\cc^{(i)}}(\mathbf{F})}\).
To increase speed, \ac{MFCI} can (optionally) use an approximate update of the harmonic flow instead.
Consider that, in sparse configurations, there are few overlaps between $2$-cells from different iterations.
Thus, the change in harmonic flow depends mostly on the boundaries \(\hat{\mathbf{b}}^{(i)}_1, \ldots, \hat{\mathbf{b}}^{(i)}_{l'}\) of the chosen cells.
With $\mathbf{H}^{(0)} = \mathbf{F}$, we can approximate the harmonic flow as:
\begin{equation}\label{eq:iterative-update:harmonic}
	\mathbf{H}^{(i)} \gets \mathbf{H}^{(i-1)} -
	\left[\hat{\mathbf{b}}^{(i)}_1 \cdots \hat{\mathbf{b}}^{(i)}_{l'} \right] \cdot
	\left[\hat{\mathbf{b}}^{(i)}_1 \cdots \hat{\mathbf{b}}^{(i)}_{l'} \right]^\dagger \cdot
	\mathbf{B}^{(i)} 
	\mathbf{C}^{(i)}
\end{equation}

\subheading{Matrix factorization.}
To find a suitable matrix factorization \(\mathbf{F} \approx \mathbf{B} \cdot \mathbf{C}\) for our application, let us first collect some desirable properties for this factorization.
First, the approximation error \(\norm{\mathbf{F - B \cdot C}}\) should be low.
To make the discretization of the columns of \(\mathbf{B}\) to 2-cells easy, the entries of \(\mathbf{B}\) should also almost fulfill the properties of a boundary matrix already:
First, the entries of \(\mathbf{B}\) should be  \(0\) or \(\pm 1\).
Second, the boundary of $\mathbf{B}$ should be zero: $\mathbf{B}_1\mathbf{B} = 0$.

The first possible factorization we consider is a (truncated) \ac{SVD} of $\mathbf{F}$, which is the optimal low-rank approximation.
However, the entries are typically not close to $0$ or \(\pm 1\).
Hence, we also consider \ac{ICA} \cite{HYVARINEN2000411}, which separates a matrix into statistically independent components that are not necessarily orthogonal.
In practice, this is often closer to the desired decomposition into a boundary matrix $\mathbf{B}_2$ and associated circulations around $2$-cells.

\subheading{Obtaining 2-cell candidates.}
We apply the chosen matrix factorization strategy iteratively to find 2-cell candidates.
In each iteration, we calculate a low-rank matrix approximation \(\mathbf{B}^{(i)} \cdot \mathbf{C}^{(i)} \approx \mathbf{H}^{(i-1)}\) (e.g.\ using the \ac{SVD}).
In each iteration, we first calculate the approximation error of each column of the matrix $\mathbf{B}^{(i)}$ as \(\|\mathbf{H}^{(i-1)} - \mathbf{B}^{(i)}_{-,j} \mathbf{C}^{(i)}_{j,-}\|_1\), where $\|A\|_1 = \sum_{i,j}|A_{i,j}|$ denotes the $L_1$ norm.
We keep the \(l\) columns \(\mathbf{b}^{(i)}_1, \ldots, \mathbf{b}^{(i)}_l\) of \(\mathbf{B}^{(i)}\) with the lowest approximation error and discretize them to valid 2-cell candidates using one of two heuristics:
\begin{itemize}
	\item The \emph{deterministic heuristic} creates a candidate by adding edges in decreasing order of their absolute values in \(\mathbf{b}^{(i)}_j\) to an empty graph until a (unique) cycle is formed.
	\item The \emph{random-walk} based heuristic simulates a random walker, using the corresponding values from chosen vectors of the matrix factorization \(\mathbf{b}^{(i)}_j\)
	      as weights for the probability distribution.
	      Once the random walk forms a cycle, this cycle is returned as the candidate.
	      Each cycle identifies a 2-cell \(\theta'^{(i)}_j\).
\end{itemize}

We add the $l'$ best candidates to the cell complex (see \cref{eq:iterative-update}); for $l=l'$, we skip the evaluation.

Our approach infers a 2-cell candidate for each column of the approximation.
The number of 2-cell additions in each iteration thus depends on the number of chosen columns,
and the rank of the matrix approximation.
Since the rank of $\mathbf{F}$ is bounded by $s$, multiple cells can only be added for $s > 1$.
As with a larger number of samples $s$, the low-rank approximation is more likely to eliminate noise in the observed flow data, we expect \ac{MFCI} to be most useful when many flows are observed.

\section{Experiments}\label{sec:experiments}
For the empirical evaluation, we compare both the approximation error and the compute time of our new approach to the state-of-the-art method \ac{SPH}\footnote{Unless stated otherwise, we use the \emph{similarity} \ac{SPH} heuristic with 11 clusters and 11 cell candidates like in the original paper. In our experiments, a larger number did not significantly improve accuracy.} \cite{hoppe2023representing} and the non-discretized \ac{SVD}, which is the optimal continuous solution.
We choose the rank of the matrix factorization to be equal to the number of candidates $l$.

We generate synthetic random \CC{}s using the algorithm from \cite{hoppe2024random}.
We sample flow signals $\mathbf{f}_i = \mathbf{B}_2 c_i + \mathbf{f}'_i$, where $c_i \in C_2$ is a signal obtained from an i.i.d.\ gaussian distribution on all $2$-cells and $\mathbf{f}'_i \in C_1$ is i.i.d.\ gaussian noise on the edges.

We also consider real-world data based on taxi trips in New York City \cite{benson-gleich-lim,chriswhong}.
The transitions between neighborhoods in a driver's trajectory are represented as a directed flow.
These flows are then aggregated into $s$ flows via summation.

\subheading{Synthetic Data.}
\Cref{fig:lr-comp-on-er} shows the approximation errors and compute times on synthetic data.
As the \ac{ICA}-based factorization approach outperforms the \ac{SVD}-based approach in both approximation error and compute time, we only show the former.
For the most significant speed-up, we skip evaluation of candidates and use the approximate update of the harmonic flow.
With this configuration (denoted \enquote{8oo-1} in \cref{fig:lr-comp-on-er}), no calls to LSMR are required.
Thus, the computation is fast while the approximation error is comparable to \ac{SPH}.
By evaluating candidates and only adding the best in each iteration (like \ac{SPH} does), \ac{MFCI} outperforms \ac{SPH} in terms of approximation error, with a similar (if slightly larger) computational requirement.
In fact, the approximation error is close to the ground truth of the sampled cells.

\newcommand{\figheight}{0.175\textheight}
\begin{figure}[t]
	\includegraphics[height=\figheight]{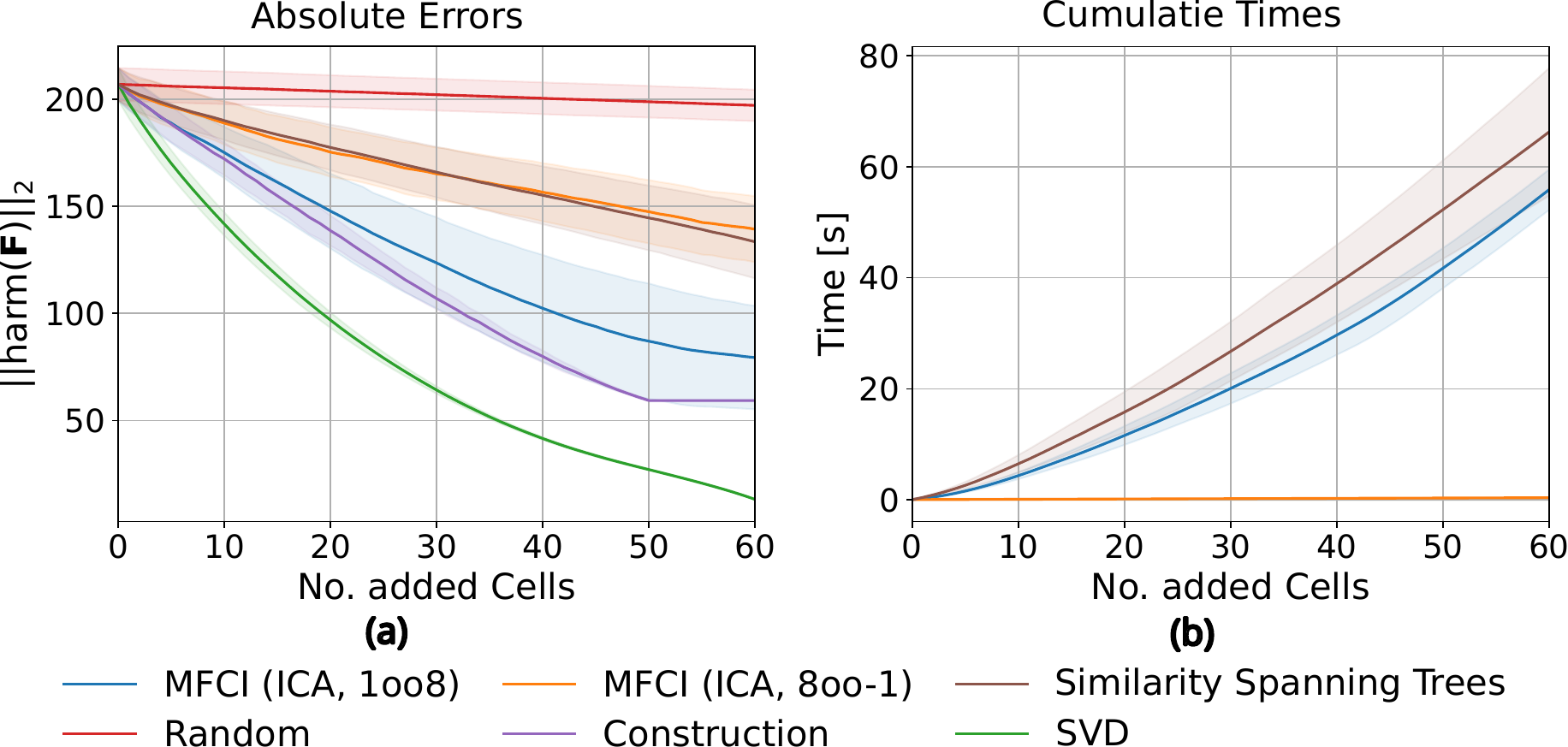}
	\vspace{-5pt}
	\caption{ \ac{MFCI} with deterministic candidate heuristic and \ac{SPH} on an Erdős–Rényi with
		\(n=40\), \(p=0.9\), 50 sampled 2-cells, 64 flows and edge noise with \(\sigma = 0.3\).
		\enquote{1oo8} means that the best cell of 8 candidates got added each iteration
		and \enquote{8oo-1} adds all 8 cells that get inferred each iteration.
		In \textbf{(a)} the SVD is the theoretical mathematical optimum.
		In \textbf{(b)} SVD, Construction and Random have no corresponding time, because they are only shown for reference.
	}\label{fig:lr-comp-on-er}
\end{figure}

\begin{figure}[t]
	\begin{center}
		\includegraphics[height=\figheight]{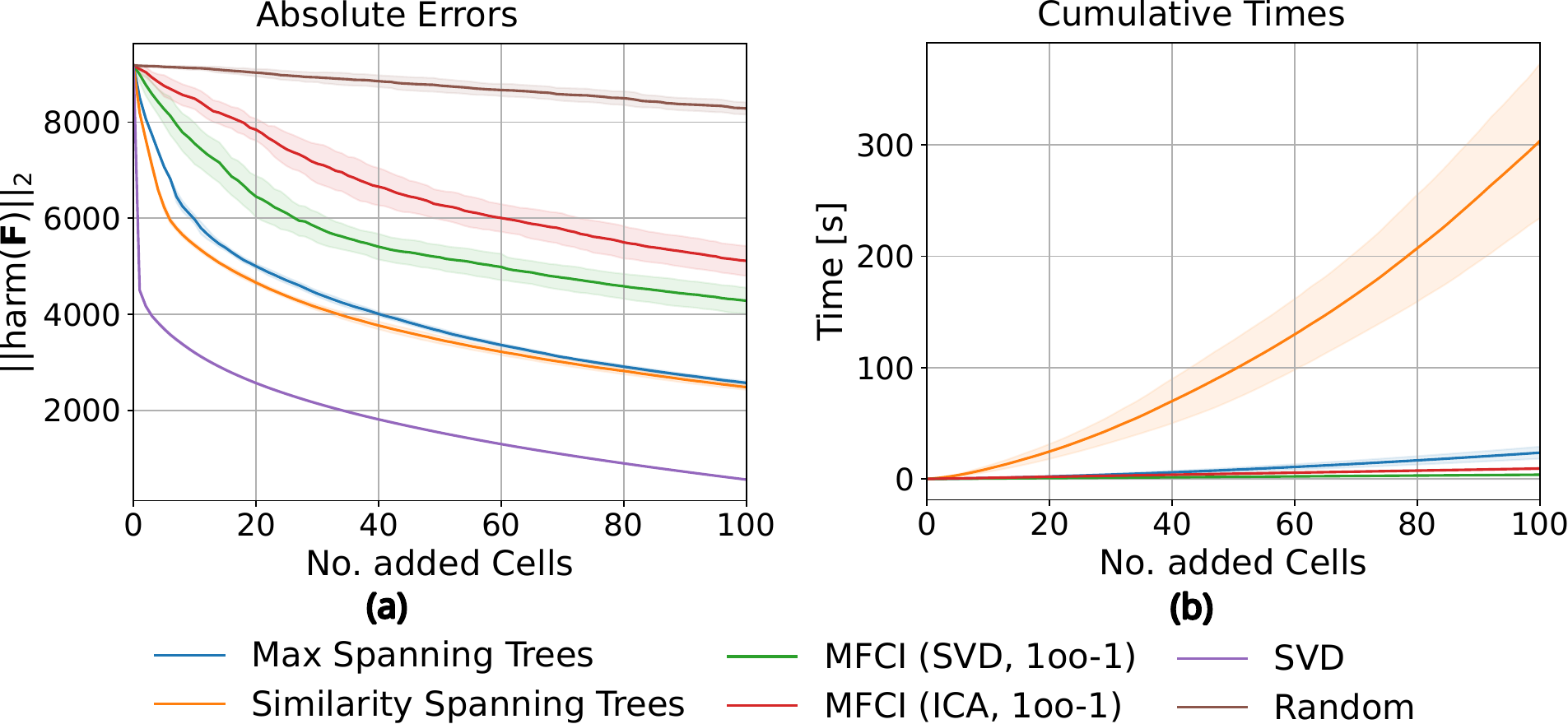}
	\end{center}
	\vspace{-10pt}
	\caption{
		\ac{MFCI} (random walk heuristic) vs \ac{SPH} on Taxi Set with 128 flows. The Max Spanning Trees only evaluates 1 candidate per iteration.
		The \ac{SVD} shows the approximation loss of \(\mathbf{F}\) using the mathematical optimal approximation
		without constraints on the matrices as required by problem \ref{eq:problem-variant-matfact}.
		In subfigure \textbf{(a)}, we see the size of the projected harmonic flow depending on the added 2-cells.
		In subfigure \textbf{(b)}, the cumulative times are presented.
	}
	\label{fig:taxi-tradeoffs}
\end{figure}

\subheading{Real-World Data.}
In \cref{fig:taxi-tradeoffs}, we compare \ac{MFCI} to \ac{SPH} on the taxi dataset.
The maximum spanning tree heuristic is configured to only evaluate one candidate per iteration, which requires only one call to LSMR.

In contrast to the synthetic data, the \ac{SPH} heuristics are more accurate than \ac{MFCI}.
Furthermore, in \ac{MFCI}, using the \ac{SVD} significantly outperforms the \ac{ICA}, which is just slightly better than the random baseline.
However, it is also notable that, after 40 added 2-cells, the difference in error between \ac{SPH} and \ac{MFCI}
stays approximately constant.
The computational times behave similarly to the synthetic data, with \ac{MFCI} (without candidate evaluation) being significantly faster than \ac{SPH}.

\begin{figure}[t]
	\begin{center}
		\vspace{-4pt}
		\includegraphics[height=\figheight]{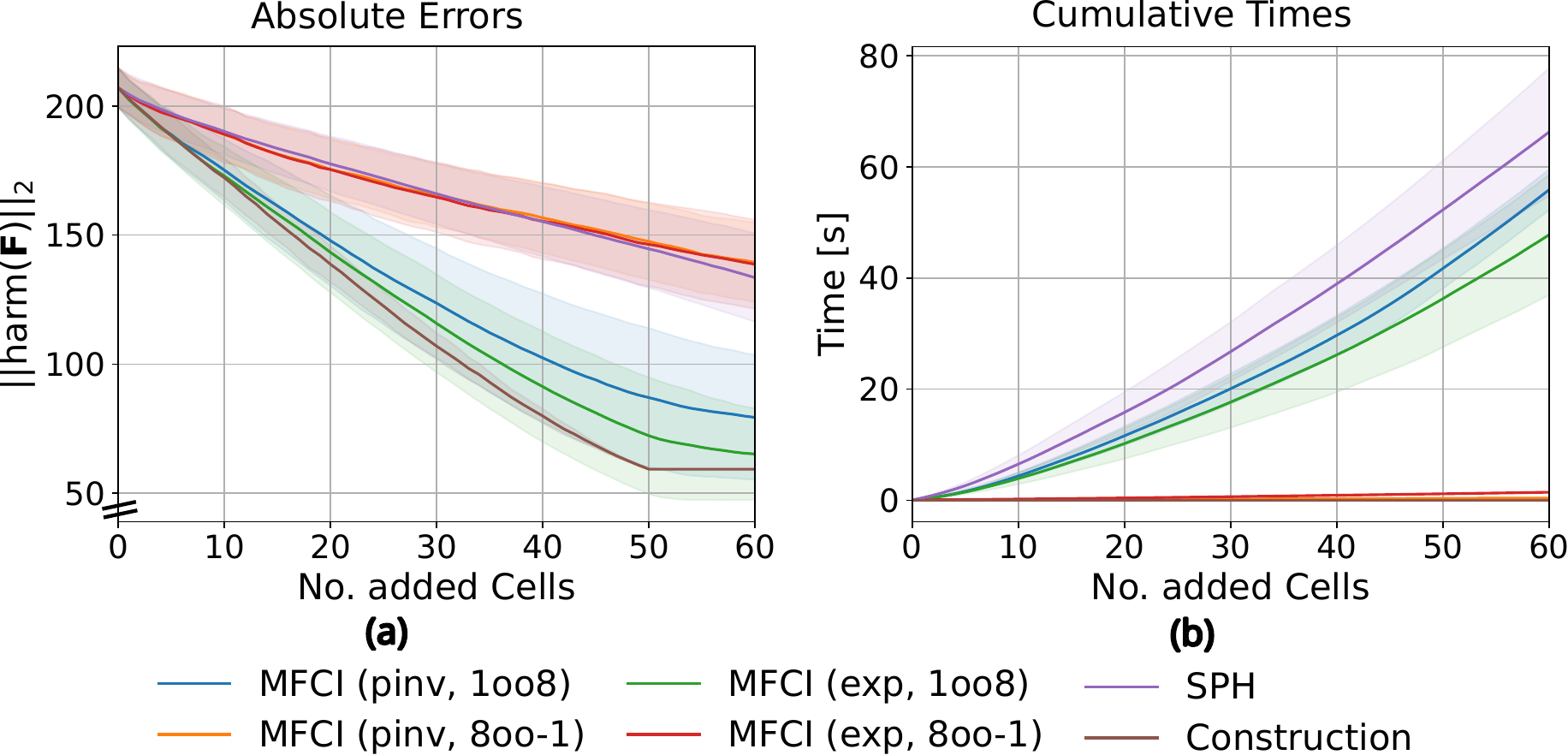}
	\end{center}
	\vspace{-10pt}
	\caption{Comparison of using explicit LSMR projection (exp) and the approximation via pseudoinverse (pinv) to calculate the remaining harmonic flow. Candidates obtained using the deterministic heuristic.}
	\label{fig:pinv-vs-exp}
	\vspace{-3pt}
\end{figure}

\subheading{Harmonic Flow Approximation.}\label{sec:sub:harmonic-flow-approx}
We evaluate the usefulness of the proposed approximate calculation of the harmonic flow using the pseudoinverse.
\Cref{fig:pinv-vs-exp} compares the configurations of \ac{MFCI} from \cref{fig:lr-comp-on-er} to the equivalent approximation variants.
For \ac{MFCI} with candidate evaluation, we see a small increase in approximation error with a negligible performance improvement when using the approximation.
However, for the faster variant without candidate evaluation, the error does not increase, but the computation time is less than half of the original.
Overall, this suggests that the approximation is a good trade-off for the faster variant of \ac{MFCI}.

\begin{figure}
	\vspace{-3pt}
	\begin{center}
		\includegraphics[height=\figheight]{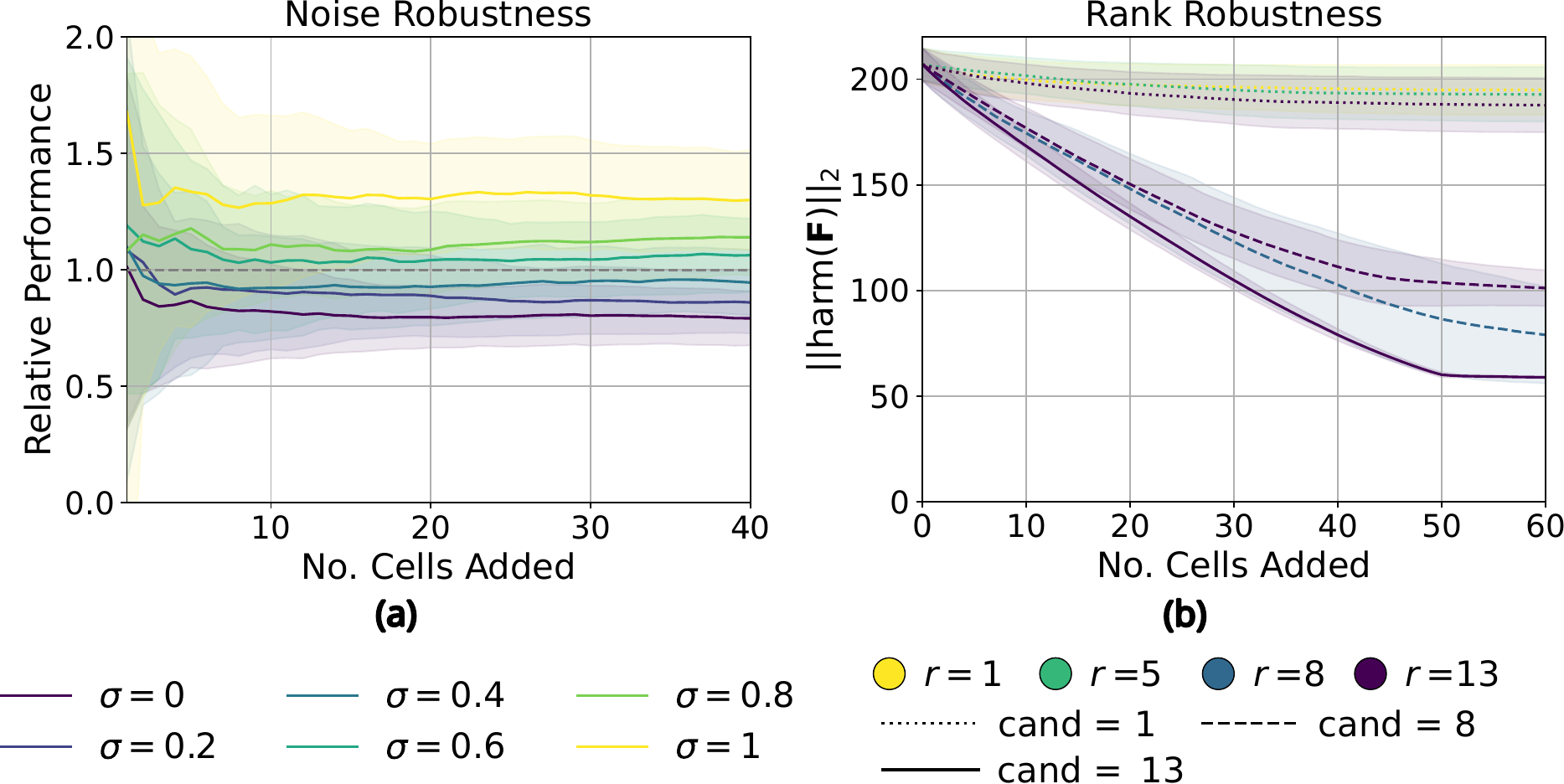}
	\end{center}

	\vspace{-10pt}
	\caption{\textbf{(a)} Relative performance of \ac{MFCI} (SVD, best candidate out of $5$, approx.\ calculation of harmonic flow, deterministic candidate heuristic) compared to \ac{SPH} on a \CC{} sampled from Erdős–Rényi with \(n=40, p=0.9\), 80 sampled 2-cells, and 64 flows.
		The flow and the noise were drawn from normal distributions with mean $\mu=0$; the former with a standard deviation \(\sigma = 1\) and the latter according to the legend.
		We show the relative performance because the inferred cells and approximation error change with the noise level.
		The relative performance is calculated as \((r - a)/(r-b)\) where \(r\) is the average error of a random algorithm, \(a\) is the error of \ac{MFCI} and \(b\) is the error of \ac{SPH}.
		As such, a relative performance of $0$ is as good as adding random cells; a value of $1$ is as good as \ac{SPH}.
		A value above $1$ indicates that \ac{MFCI} outperforms \ac{SPH}.
    \textbf{(b)} Different no. of ranks and candidates for MFCI (ICA, 1oo\_); showing negligible impact of rank on accuracy.  Experiment settings from \cref{fig:lr-comp-on-er}; \looseness=-1
	}\label{fig:noise_robustness}
\end{figure}

\subheading{Robustness to Noise.}
\Cref{fig:noise_robustness}a shows the relative performance of \ac{MFCI} compared to \ac{SPH} for different noise levels.
We observe that for low noise levels, \ac{SPH} is more accurate than \ac{MFCI}, but the opposite is true for higher noise levels.
We hypothesize that the low-rank matrix factorization  filters out noise in the data, leading to better candidate cells than in \ac{SPH}.\looseness=-1

\section{Conclusion}
We presented a new framework, \ac{MFCI}, that approximately solves the general cell inference problem by using matrix factorization.
Overall, our experiments show that compared to the previous state-of-the-art, \ac{MFCI} achieves a significant speed-up at the expense of a small increase in the approximation error in certain settings.
Furthermore, in noisy configurations, \ac{MFCI} achieves a better approximation error.

Importantly, we saw qualitatively different results on synthetic and real-world data.
The synthetic data was based on the same assumptions made in the development of \ac{MFCI}, namely, that there are underlying $2$-cells that generate flow \emph{independently} of each other.
It is likely that real-world data has strong correlations, leading to a less useful matrix decomposition.
This hypothesis is consistent with the observation that \ac{ICA}, which employs similar inherent assumptions, performs poorly on real-world data.
Nevertheless, \ac{MFCI} with \ac{SVD} is useful for applications where its significant speed-up (and thus increased scalability) outweighs its relatively small increase in approximation error.
The approximation error is based on signal compression and likely translates well to signal processing tasks, but it is unclear how it affects the per\-formance of other downstream tasks built on the inferred cells.

There is a large space of possible configurations for \ac{MFCI} that exceeds the scope of this paper.
First, any component of \ac{MFCI} could be replaced, mainly the matrix factorization method and the heuristic to obtain candidates.
Furthermore, the projection could use a hybrid approach where the approximate method is used, but the projection is done explicitly in some iterations.
Second, due to the iterative nature of both \ac{MFCI} and \ac{SPH}, it is also possible to combine the two methods.
As we have seen, \ac{SPH} performs particularly well in the first iterations, where it is also comparatively fast.
Therefore, it may be advantageous to design a hybrid approach, where the first few iterations are performed with \ac{SPH} and the remaining iterations with \ac{MFCI}.
Ideally, this could combine the accuracy of \ac{SPH} in the early iterations with the better computational performance of \ac{MFCI} in the later, more computationally expensive, iterations.

\bibliographystyle{IEEEtran}
\bibliography{IEEEabrv,reference.bib}

\end{document}